\documentclass{llncs}

\usepackage{amsmath}
\usepackage{amsfonts}
\usepackage{graphicx}
\usepackage{algorithm}
\usepackage{algpseudocode}
\usepackage{caption}
\usepackage{tikz}
\usetikzlibrary{patterns}
\graphicspath{{images/}}
\begin{document}
\title{On the Effects of Measurement Uncertainty in Optimal Control of Contact Interactions}
\author{Brahayam Pont\'on\inst{1} \and Stefan Schaal\inst{1} \inst{2} \and Ludovic Righetti \inst{1}}
\institute{Max-Planck Institute for Intelligent Systems, Tuebingen-Germany
			\email{firstname.lastname@tuebingen.mpg.de}
           \and University of Southern California, Los Angeles-USA \\
           \email{sschaal@usc.edu}}
\maketitle
%
\begin{abstract}
Stochastic Optimal Control (SOC) typically considers noise only in the process model, i.e. unknown disturbances. However, in many robotic applications involving interaction with the environment, such as locomotion and manipulation, uncertainty also comes from lack of precise knowledge of the world, which is not an actual disturbance. We analyze the effects of also considering noise in the measurement model, by developing a SOC algorithm based on risk-sensitive control, that includes the dynamics of an observer in such a way that the control law explicitly depends on the current measurement uncertainty. In simulation results on a simple 2D manipulator, we have observed that measurement uncertainty leads to low impedance behaviors, a result in contrast with the effects of process noise that creates stiff behaviors. This suggests that taking into account measurement uncertainty could be a potentially very interesting way to approach problems involving uncertain contact interactions.
\end{abstract}
%
\section{Introduction}
\vspace{-0.2cm}
In a not distant future, personal robots will be a common part of our daily lives, with a broad range of applications going from industrial and service applications to common household scenarios \cite{learningcontrol}. Being able to safely operate among humans by \emph{optimally adapting to uncertainty} in a dynamic environment is a key ingredient for this to happen. In this contribution, we address this aspect by studying the effects of measurement uncertainty in stochastic optimal control problems.

\begin{figure}[h]
	\centering
	\includegraphics[width=90mm]{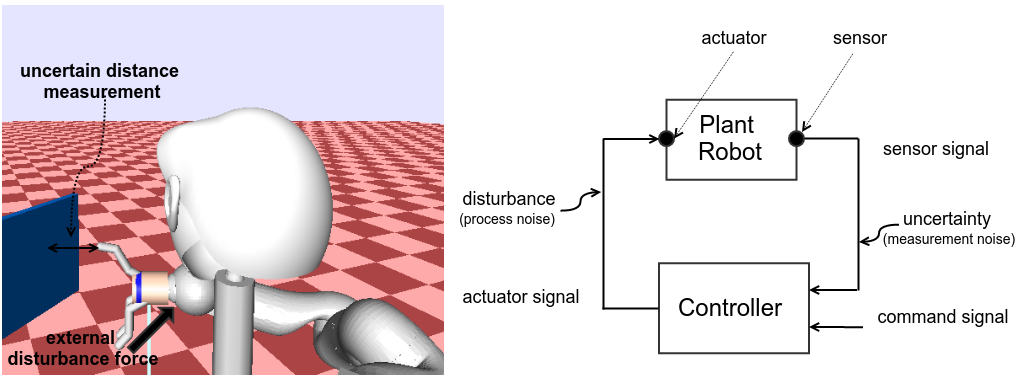}
	\caption{\textit{To the left}: Schematic of two different sources of uncertainty. \textit{To the right}: Simplified control diagram showing where and how they enter the system.}
	\label{fig_apollo}
	\vspace{-0.6cm}
\end{figure}

For instance, we would like to understand the effects of considering uncertainty information upon optimal control solutions in problems that involve contact interactions. We distinguish between two sources of uncertainty: one due to external forces that physically perturb the robot, and the other due to uncertain knowledge of the robot’s state, inferred from noisy measurements. Figure \ref{fig_apollo} shows a schematic. On the one hand, external disturbances (process model’s noise) directly affect the dynamic evolution of the robot and a stiff behavior is usually adopted to control the robot in their presence. On the other hand, our belief about the robot state, e.g. distance to a contact location, can be thought of as a noisy sensor signal.
This, however, does not affect our actuation directly; instead, it influences the way in which decisions over control signals are taken. For example, if we are walking down a stair in the dark and we are uncertain about the floor location because we cannot see it, we reach for it with a gentle touch in order not to harm ourselves or fall. We observe that in this case, compliance was used to handle measurement uncertainty.

\vspace{-0.6cm}
\emph{Why is it important to consider measurement noise effects?} First of all, properly addressing robustness issues due to uncertainty is important in robotics \cite{learningcontrol}. 
More concretely, imagine a reaching motion: in general, increasing the magnitude of an external disturbance implies one must increase the stiffness of the robot to maintain a desired tracking accuracy. Now, consider the experiment under measurement noise, e.g. you would like to grasp an object using no vision information, or you try to reach a wall to orient yourself in a dark corridor. Under these conditions, compliance is key to carefully reach an object. This distinction is important, because it suggests that modeling interactions of a robot with its environment (a fundamental problem in robotics) as an optimal control problem with measurement uncertainty, one could naturally get optimal compliant behaviors as a function of the uncertainty level.

Optimal control techniques based on \emph{applying Bellman's Principle of Optimality around nominal trajectories}, e.g. techniques such as DDP or iLQG, have been very successfully used in robotics for large degrees of freedom system \cite{SiyuanFeng,MichaelNeunert,TassaMPC,control_limited_DDP,iLQGTodorov}. They maintain a single trajectory as a local method; and improve it iteratively based on dynamic programming along a neighborhood of the trajectory. This allows them to overcome to some extent the curse of dimensionality while remaining computationally efficient. Our algorithm belongs to this family, but distinguishes itself by the ability to explicitly consider measurement uncertainty.

Typically these methods find a solution for a nonlinear problem by iteratively solving a first- or second-order Taylor approximation of it \cite{DDPmayne,SiderisSLQ}. 
This, however, imposes limitations: it only considers the mean of the objective function (expectation of a quadratic form) and systems under purely additive noise. As a result, the optimal control for the stochastic and deterministic problems is the same, i.e. uncertainty is dealt with by the estimator separately, and the control design is independent from noise. 
 While it is reasonable for systems with small noise intensity, intuitively, one would not expect the same to be true for systems with large noise intensity, where a \emph{control strategy capable of reasoning about noise statistics and cost of uncertainty} would be more appropriate. 
 While there exists other equally valid methods for dealing with uncertainty, e.g. avoiding uncertain regions of the state space \cite{johnson2016convergent}, we are interested in understanding how to control desired contact interactions with an inherently uncertain environment. This is the reason why we prefer a risk-sensitive over a risk-neutral optimal control approach. In this way, our algorithm is sensitive to the statistical properties of the noise and can incorporate this information into the optimal control law.

Risk-sensitive optimal control algorithms usually invalidate the assumptions of the Certainty Equivalence Principle, e.g. by using a nonlinear state equation, multiplicative noise or a non-quadratic objective function. Care is taken though at preserving the computational efficiency. The problem of considering multiplicative process noise was studied in \cite{iLQGTodorov}, and extended to multiplicative measurement noise in \cite{iLQGWeiwei,Todorov2005b}. These methods construct an affine control law dependent on noise statistics and apply it to control a two-DOF model of a biomechanical arm. Another appealing alternative is to capture noise effects \emph{on higher order statistics of the objective function by using non-quadratic costs}.

Jacobson \cite{originalLEG} introduced a Linear-Exponential-Gaussian (LEG) algorithm to consider higher order statistics \emph{using as cost an exponential transformation of the original objective function}. He derived feedback controllers for a linear system with additive process noise that explicitly depend on noise statistics. At low noise levels, the LEG control law is similar to the Linear-Quadratic-Gaussian (LQG) controller; but the larger the noise, the more they differ. This idea was extended in \cite{iLEG} for continuous-time nonlinear SOC problems using an iterative algorithm and they illustrated both risk-seeking and risk-averse behaviors in a continuous-time cliff problem. A more comprehensive review on the risk-sensitive literature for systems under process noise can be found in \cite{LinearlySolvableControl}, where they present a unified theory of linearly solvable optimal control problems that includes both standard and risk-sensitive optimal control problems.

It has been shown \cite{extendedLEG}, that due to the multiplicative nature of the exponential cost, it is not straightforward to extend the results for the case of measurement noise (where the control law is not a linear functional of the current state, but of the whole smoothed history of states). As solution, they proposed to define a state that grows every timestep to comprise the entire history of states seen so far. Because of this increasingly growing computational complexity, only two cases, where simplifications occur, are practical: when the objective function is a functional only of the final state, and when there is no process noise. 

In this paper, we use recent results in risk-sensitive control \cite{iLEG} and extend them to incorporate measurement uncertainty. We will then show how different types of noise can significantly change the optimal controls. In our approach, as in \cite{iLEG}, we sequentially approximate the nonlinear problem and design risk-sensitive controllers. However, in order to include measurement uncertainty effects, instead of using a growing state composed of the entire history of states, we \emph{use an enlarged dynamical system composed of the control and estimation problems} \cite{paper_baras,paper_whittle}, where the number of states only doubles. We reduce the amount of information for constructing the optimal control to statistics that can be captured in the state estimate (i.e. expectation and variance). But by doing this, we gain increased flexibility at designing the objective function and are able to simultaneously capture process and measurement noise to compare their effects.

The most important contributions of this work are as follows:
\begin{itemize}
\item We propose a theoretical contribution, where we extend recent work on risk-sensitive control to the measurement noise case by incorporating a state observer. This makes the optimal control explicitly dependent on statistical properties of process and measurement uncertainty.
\item By applying our algorithm in a contact interaction experiment of a 2D manipulator with an uncertain wall, we show that our approach produces optimal impedance behaviors for contact interaction, that differ from the usual stiff behaviors in that compliance is encouraged under measurement noise.
\end{itemize}

In the following, we present the problem formulation and background material. Then, we show the algorithm derivation and illustrate its performance in two simple robotic tasks: a viapoint and a contact interaction task.
%
\vspace{-0.2cm}
\section{Problem Formulation and Background Material}
\vspace{-0.2cm}
In this section, we present the stochastic optimal control problem under measurement noise. The following stochastic differential equations (SDE's) define the dynamical evolution of the state and measurement models respectively
\begin{align}
d\vec{x} =& \vec{m}(\vec{x},\vec{u}) dt + \vec{M}(\vec{x},\vec{u})d\vec{\omega} \enspace , \label{eqn_dyn}  \\
d\vec{y} =& \vec{n}(\vec{x},\vec{u}) dt + \vec{N}(\vec{x},\vec{u})d\vec{\gamma} \enspace . \label{eqn_mea}
\end{align}

Let $\vec{x} \in \bbbr^{n}$, $\vec{u} \in \bbbr^{m}$ and $\vec{y} \in \bbbr^{p}$ be the system states, control and measured outputs. $d\vec{\omega}$ and $d\vec{\gamma}$ are zero-mean Brownian motions with covariance $\vec{\Omega} dt$, $\vec{\Gamma} dt$. $\vec{m}(\vec{x},\vec{u})$ and $\vec{n}(\vec{x},\vec{u})$ are the drift coefficients representing the deterministic components of the dynamics and measurement models. $\vec{M}(\vec{x},\vec{u})$ and $\vec{N}(\vec{x},\vec{u})$ are the diffusion coefficients that encode the stochasticity of the problem.

In optimal control, we are interested in minimizing an objective function $\mathcal{J}^{\vec{\pi}}$, which is a functional of the control policy $\vec{u}=\vec{\pi}(\vec{x})$, defined as
\begin{equation}
\mathcal{J}^{\vec{\pi}}(\vec{x},t) = \Phi_{f}(\vec{x}_{t_{f}})+\int_{t}^{t_{f}}L(\vec{x}_t,\vec{u}_t,t)dt \enspace , \label{performance_index}
\end{equation}
\noindent where $L(\vec{x}_t,\vec{u}_t,t)$ is the rate at which cost increases. $\Phi_{f}(\vec{x}_{t_{f}})$ is the cost at the final time $t_{f}$. In standard optimal control, the mean of the objective function $\mathbb{E}[\mathcal{J}^{\vec{\pi}}]$ would typically be minimized. However, in order to analyze the effects of uncertainty on the optimal controls, this is not sufficient. We need to include the notion of the cost of uncertainty into the objective function. For this purpose, we use tools from the risk-sensitive control literature, that allow us to include higher order statistics of the objective function $\mathcal{J}^{\vec{\pi}}$ in the minimization. This is done by reformulating the objective function as an exponential transformation of the original objective function \cite{originalLEG}. The risk-sensitive cost is then given by
\begin{equation}
J = \min_{\vec{\pi}} \mathbb{E} \{\exp[\sigma \mathcal{J}^{\vec{\pi}}]\} \enspace .
\label{general_cost}
\end{equation}
$\mathcal{J}^{\vec{\pi}}$ is a random variable functional of the policy $\vec{u}=\vec{\pi}(\vec{x})$. $\sigma \in \bbbr$ is the risk-sensitive parameter, $\mathbb{E}$ is the expectation over $\mathcal{J}^{\vec{\pi}}$. $J$ is therefore the risk-sensitive cost and corresponds to the moment generating function, an alternative specification of the probability distribution of the random variable $\mathcal{J}^{\vec{\pi}}$ \cite{PathIntegralsForStochasticProcesses}.

In the following, we recall two previous results elaborated in \cite{iLEG,originalLEG,extendedLEG} that we will use for the development of our approach: the meaning of the transformed cost and the form of the Hamilton Jacobi Bellman (HJB) equation under the exponential transformation.
\vspace{-0.2cm}
\subsection{Meaning of the Exponential Transformation of the Cost}
It has been shown \cite{iLEG} that the cumulant generating function (logarithmic transformation of the moment generating function) of the risk-sensitive cost $J$ can be rewritten as a linear combination of the moments of the objective function $\mathcal{J}^{\vec{\pi}}$
\begin{equation}
\frac{1}{\sigma}\log{[J]} = \mathbb{E}[\mathcal{J}^{\vec{\pi}}] + \frac{\sigma}{2} {\mu}_{2}[\mathcal{J}^{\vec{\pi}}] + \frac{\sigma^{2}}{6} {\mu}_{3}[\mathcal{J}^{\vec{\pi}}] + \cdots \enspace ,
\label{cgf}
\end{equation}
${\mu}_{2}$, ${\mu}_{3}$ denote the variance and skewness of $\mathcal{J}^{\vec{\pi}}$. Therefore, the risk-sensitive cost is a linear combination of all the moments of the original objective function. It provides an additional degree of freedom, namely the risk-sensitive parameter $\sigma$, which allows us to define if the higher order moments act as a penalty or a reward in the cost, giving rise to risk-averse or risk-seeking behaviors respectively. Depending on $\sigma$, there will be a compromise between increasing control effort and narrowing confidence intervals. The lower the values of $\sigma$, the less weight is given to higher order moments. When it is negative, they even act as a reward, therefore leading to risk-seeking solutions.
\vspace{-0.2cm}
\subsection{HJB Equation under the Exponential Transformation}
From \cite{iLEG}, we recall the form of the HJB equation under the exponential transformation (the dynamics are given only by \eqref{eqn_dyn} and the cost by \eqref{performance_index}-\eqref{general_cost})
\begin{equation}
	\underbrace{-\partial_{t} \Psi = \min_{u}
	\bigg\{L + \nabla_{\mathbf{x}} \Psi^{T} \mathbf{m} + \frac{1}{2} {Tr} \left(\nabla_{\mathbf{x} \mathbf{x}} \Psi  \mathbf{M} \mathbf{M}^{T} \right)}_{\text{Usual HJB equation}} + \underbrace{ {\frac{\sigma}{2} \nabla_{\mathbf{x}} \Psi^{T}  \mathbf{M} \mathbf{M}^{T} \nabla_{\mathbf{x}} \Psi \color{black} \bigg\}} }_{\text{Term due to uncertainty}} \, ,
\label{hjb_exp}
\end{equation}
where the value function $\Psi$ is a function of $\vec{x}$ and $t$.
The terms without $\sigma$ represent the usual HJB equation for a stochastic dynamical system with cost rate $L$ due to the current state and control, the free drift and control benefit costs, and the diffusion cost. The interesting term is the last one which captures noise effects on statistical properties of the cost (higher moments). When $\sigma$ is zero, the problem reduces to the minimization of the usual expected value of the cost $E[\mathcal{J}^{\vec{\pi}}]$. It is worth highlighting that these two results presented as background material model only process noise and do not include measurement noise.
\vspace{-0.2cm}
\subsection{Problem Formulation}
Finally, we conclude the problem definition. The cost to minimize is given by \eqref{performance_index}-\eqref{general_cost}. Our goal is to find a risk-sensitive optimal control law $\vec{\pi}^{*}$ that minimizes the cost $J^{\vec{\pi}}(\vec{x}_{0},t_{0})$ for the stochastic system in the presence of additive process and measurement noise \eqref{eqn_dyn}-\eqref{eqn_mea}. The globally optimal control law $\vec{\pi}^{*}(\vec{x},t)$ does not depend on an initial state. However, finding it is in general intractable. Instead, we are interested in a locally-optimal feedback control law that approximates the globally optimal solution in the vicinity of a nominal trajectory $\vec{x}^{n}_{t}$. Since this nominal trajectory depends on the initial state of the system, so does the optimal feedback control law. As can be noted, our formulation of the problem differs from the results in the background material, because we include measurement noise. However, as we will show in detail in the next section, these results can still be used in our case, after a certain reformulation of the problem.
\vspace{-0.2cm}
\section{Algorithm Derivation}
\vspace{-0.2cm}
As mentioned in the Introduction, there have been numerous contributions on risk-sensitive control with process noise \cite{LinearlySolvableControl,iLEG}. Therefore, our goal in this section is to reformulate our problem including the stochastic dynamics of the measurement model, such that we can use some of these previous results. The algorithmic idea is to extend the state dynamics \eqref{eqn_dyn}, with the dynamics of a state estimator. As will be seen later in detail, the key element then is to include a forward propagation of measurement uncertainty along a nominal trajectory and to precompute optimal estimation gains. This allows for the use of standard techniques to compute backwards in time optimal feedback controllers \cite{DDPmayne}. This idea, however, is not particular to the algorithm we present in the following, but could be used in combination with other methods, such as the ones presented in \cite{LinearlySolvableControl}. In the following, we derive a continuous time algorithm\footnote{The derivation of the discrete-time version of the algorithm is presented in the accompanying appendix.}.

At each iteration, the algorithm begins with a nominal control sequence $\vec{u}^{n}_{t}$ and the corresponding zero-noise trajectory $\vec{x}^{n}_{t}$, obtained by applying the control sequence to the dynamics $\vec{\dot{x}} =  \vec{m}(\vec{x},\vec{u})$ with initial state $\vec{x}(0)=\vec{x}_{0}$. Next, we follow a standard approach in iterative optimal control \cite{DDPmayne,SiderisSLQ} to form a linear approximation of the dynamics and a quadratic approximation of the cost along the nominal trajectories $\vec{x}^{n}_{t}$ and $\vec{u}^{n}_{t}$, in terms of state and control deviations $\delta \vec{x}_{t} = \vec{x}_{t} - \vec{x}^{n}_{t}$, $\delta \vec{u}_{t} = \vec{u}_{t} - \vec{u}^{n}_{t}$. Dynamics and measurement models then become
\begin{align}
d(\delta \vec{x}_{t}) =& (\vec{A}_{t} {\delta \vec{x}_{t}} + \vec{B}_{t} {\delta \vec{u}_{t}}) dt + \vec{C}_{t} d\vec{\omega}_{t} \enspace , \label{lin_dyn_CT} \\
d(\delta \vec{y}_{t}) =& (\vec{F}_{t} {\delta \vec{x}_{t}} + \vec{E}_{t} {\delta \vec{u}_{t}}) dt + \vec{D}_{t} d\vec{\gamma}_{t} \enspace . \label{lin_mea_CT}
\end{align}
Evaluated along the nominal trajectories $\vec{x}^{n}_{t}$, $\vec{u}^{n}_{t}$, the matrices are given by
\begin{align*}
&\vec{A}_{t} = {\partial \vec{m}(\vec{x},\vec{u})}/{\partial \vec{x}^{T}} |_{\vec{x}^{n}_{t}, \vec{u}^{n}_{t}}, \, 
\vec{B}_{t} = {\partial \vec{m}(\vec{x},\vec{u})}/{\partial \vec{u}^{T}} |_{\vec{x}^{n}_{t}, \vec{u}^{n}_{t}}, \,
\vec{C}_{t} = \vec{M}(\vec{x}^{n}_{t}, \vec{u}^{n}_{t}) \enspace . \\
&\vec{F}_{t} = {\partial \vec{n}(\vec{x},\vec{u})}/{\partial \vec{x}^{T}} |_{\vec{x}^{n}_{t}, \vec{u}^{n}_{t}}, \,
\vec{E}_{t} = {\partial \vec{n}(\vec{x},\vec{u})}/{\partial \vec{u}^{T}} |_{\vec{x}^{n}_{t}, \vec{u}^{n}_{t}}, \,
\vec{D}_{t} = \vec{N}(\vec{x}^{n}_{t}, \vec{u}^{n}_{t}) \enspace . 
\end{align*}
In the same way, the quadratic approximation of the cost $\mathcal{J}$ is given by
\begin{align}
\tilde{\ell}(\vec{x},\vec{u}) =& q_{t} + \vec{q}_{t}^{T} \delta \vec{x}_{t} + \vec{r}_{t}^{T} \delta \vec{u}_{t} + \frac{1}{2} \delta \vec{x}_{t}^{T}  \vec{Q}_{t} \delta \vec{x}_{t} + \delta \vec{x}_{t}^{T}  \vec{P}_{t} \delta \vec{u}_{t} + \frac{1}{2} \delta \vec{u}_{t}^{T}  \vec{R}_{t} \delta \vec{u}_{t} \, , \label{running_cst_CT} \\
\tilde{\ell}_{f}(\vec{x}) =& q_{f} + \vec{q}_{f}^{T} \delta \vec{x}_{t} + \frac{1}{2} \delta \vec{x}_{t}^{T}  \vec{Q}_{f} \delta \vec{x}_{t} \enspace . \label{final_cst_CT}
\end{align}
In order to explicitly take into account noise present in our measurement model \eqref{eqn_mea}, we include the dynamics of a state observer. This is an important step,  that allows us to define an enlarged dynamical system composed of the control and estimation problems. By using this enlarged dynamical system, we are able to include measurement noise and extend previous results, while remaining computationally efficient. The state observer could in principle be of any type, but it is required that out of it, we can obtain a sequence of estimation gains. Therefore, we use an Extended Kalman filter (EKF), whose dynamics are given by
\begin{align}
d(\delta \vec{\hat{x}}_{t}) = &(\vec{A}_{t} {\delta \vec{\hat{x}}_{t}} + \vec{B}_{t} {\delta \vec{u}_{t}}) dt + \vec{K}_{t} [ d(\delta \vec{y}_{t}) - d(\delta \vec{\hat{y}}_{t})] \enspace .
\label{est_dyn_CT}
\end{align}
The dynamics of the control-estimation problem can be compactly written as
\begin{align}
\underbrace{
	\begin{bmatrix}
	d(\delta \vec{x}_{t})  \\
	d(\delta \vec{\hat{x}}_{t})
	\end{bmatrix}
}_{d(\delta \vec{\tilde{x}}_{t})} = &
\underbrace{
	\begin{bmatrix}
	\vec{A}_{t} {\delta \vec{x}_{t}} + \vec{B}_{t} {\delta \vec{u}_{t}}  \\
	\vec{A}_{t} {\delta \vec{\hat{x}}_{t}} + \vec{B}_{t} {\delta \vec{u}_{t}} + \vec{K}_{t} \vec{F}_{t}(\delta \vec{x}_{t} - \delta \vec{\hat{x}}_{t})
	\end{bmatrix}
}_{\vec{f}(\delta \vec{\tilde{x}}_{t}, \delta \vec{u}_{t} )} dt + 
&\underbrace{
	\begin{bmatrix}
	\vec{C}_{t} & 0             			   \\
	0			& \vec{K}_{t} \vec{D}_{t}
	\end{bmatrix}
}_{\vec{g(t)}}
\begin{bmatrix}
d\vec{\omega_{t}} \\
d\vec{\gamma_{t}}
\end{bmatrix} \enspace . \label{enlarged_dynamics_system_CT}
\end{align}
$\delta \vec{\hat{x}}_{t}$ is the estimate of $\delta \vec{x}_{t}$, and $\delta \vec{\tilde{x}}_{t}$ represents the vector $[\delta \vec{x}_{t},\delta \vec{\hat{x}}_{t}]^{T}$. Equation \eqref{enlarged_dynamics_system_CT} is a bilinear system in $\delta \vec{\tilde{x}}_{t}$, $\delta \vec{u}_{t}$ and $\vec{K}_{t}$. Below, we show in detail the derivation. However, the algorithm's main idea is to use this special problem structure to iteratively find a solution. We forward propagate measurement noise and compute estimation gains $\vec{K}_{t}$ along the nominal trajectory. Then, with fixed estimation gains, we use a usual  backward pass to compute feedback controllers \cite{DDPmayne,SiderisSLQ}. This eases the design of a locally optimal estimator and controller, while still being able to consider the effects of process and measurement noise. As can be easily noticed, $\vec{f}(\delta \vec{\tilde{x}}_{t}, \delta \vec{u}_{t} )$ and $\vec{g(t)}$ correspond to what in \eqref{hjb_exp}, we called $\vec{m}$ and $\vec{M}$ respectively. However, now they include measurement noise by incorporating the dynamics of a state estimator.
\vspace{-0.1cm}
\subsubsection{Estimator Design.}
We use an EKF; however, other estimators could be used as long as we can extract a sequence of estimation gains. The optimal estimation gains that minimize the error dynamics
\vspace{-0.2cm}
\begin{align}
\vec{\dot{\Sigma}}_{t}^{e} = & (\vec{A}_{t} - \vec{K}_{t} \vec{F}_{t}) \vec{\Sigma}_{t}^{e} + \vec{\Sigma}_{t}^{e} (\vec{A}_{t} - \vec{K}_{t} \vec{F}_{t})^{T} + \vec{K}_{t} \vec{D}_{t} \vec{\Gamma}_{t} \vec{D}_{t}^{T} \vec{K}_{t}^{T} + \vec{C}_{t} \vec{\Omega}_{t} \vec{C}_{t}^{T} \label{covariance_CT}
\end{align}
\vspace{-0.2cm}
are given by
\vspace{-0.2cm}
\begin{align}
\vec{K}_{t} = & \vec{\Sigma}_{t}^{e} \vec{F}_{t}^{T} (\vec{D}_{t} \vec{\Gamma}_{t} \vec{D}_{t}^{T})^{-1} \enspace . \label{optimal_estimation_gain_CT}
\end{align}

They are updated at each iteration in a forward pass along the nominal trajectories, and are then fixed for the backward pass. In this way, the estimation-control system \eqref{enlarged_dynamics_system_CT}, is linear in $\delta \vec{\tilde{x}}_{t}$ and $\delta \vec{u}_{t}$. This allows us to make use of the HJB Eq. \eqref{hjb_exp} to compute a control law $\vec{\pi}$ sensitive to both process and measurement noise of the original system.
\vspace{-0.5cm}
\subsubsection{Controller Design.}
The locally-optimal control law is affine, of the form $\delta \vec{u}_{t} = \vec{l}_{t} + \vec{L}_{t} \delta \vec{\hat{x}}_{t}$. Notice that, we assume it to be a functional only of the state estimate. The HJB equation for this system has the same form as \eqref{hjb_exp} (remember that $\vec{m}$ and $\vec{M}$ correspond now to $\vec{f}$ and $\vec{g}$ respectively), the cost is given by  \eqref{running_cst_CT}-\eqref{final_cst_CT} (remember that we use the HJB equation under the exponential transformation; therefore, the cost need not to be exponentiated), and the dynamics by \eqref{enlarged_dynamics_system_CT}. The Ansatz for the value function $\Psi(\delta \vec{\tilde{x}}_{t},t)$ is quadratic of the form
\begin{equation}
\Psi(\delta \vec{\tilde{x}}_{t},t) = \frac{1}{2}
\begin{bmatrix}
\delta \vec{x}_{t} \\ \delta \vec{\hat{x}}_{t}
\end{bmatrix}^{T}
\begin{bmatrix}
\vec{S}^{x}_{t} 			    & \vec{S}^{x \hat{x}}_{t} \\
(\vec{S}^{x \hat{x}}_{t})^{T} 	& \vec{S}^{\hat{x}}_{t}
\end{bmatrix}
\begin{bmatrix}
\delta \vec{x}_{t} \\ \delta \vec{\hat{x}}_{t}
\end{bmatrix} + 
\begin{bmatrix}
\delta \vec{x}_{t} \\ \delta \vec{\hat{x}}_{t}
\end{bmatrix}^{T}
\begin{bmatrix}
\vec{s}^{x}_{t} \\ \vec{s}^{\hat{x}}_{t}
\end{bmatrix} + s_{t} \enspace . \nonumber \label{quadratic_ansatz_CT}
\end{equation}
\noindent and the partial derivatives of the Ansatz $\Psi$ are given by
\begin{align*}
&\partial_{t} \Psi = \frac{1}{2}
\begin{bmatrix}
\delta \vec{x}_{t} \\ \delta \vec{\hat{x}}_{t}
\end{bmatrix}^{T}
\begin{bmatrix}
\vec{\dot{S}}^{x}_{t} 			& \vec{\dot{S}}^{x \hat{x}}_{t} \\
\vec{\dot{S}}^{\hat{x} x}_{t} 	& \vec{\dot{S}}^{\hat{x}}_{t}
\end{bmatrix}
\begin{bmatrix}
\delta \vec{x}_{t} \\ \delta \vec{\hat{x}}_{t}
\end{bmatrix} + 
\begin{bmatrix}
\delta \vec{x}_{t} \\ \delta \vec{\hat{x}}_{t}
\end{bmatrix}^{T}
\begin{bmatrix}
\vec{\dot{s}}^{x}_{t} \\ \vec{\dot{s}}^{\hat{x}}_{t}
\end{bmatrix} + \dot{s}_{t} \enspace . \\
&\nabla_{\delta \vec{\tilde{x}}} \Psi = 
\begin{bmatrix}
\vec{S}^{x}_{t} 			& \vec{S}^{x \hat{x}}_{t} \\
\vec{S}^{\hat{x} x}_{t} 	& \vec{S}^{\hat{x}}_{t}
\end{bmatrix}
\begin{bmatrix}
\delta \vec{x}_{t} \\ \delta \vec{\hat{x}}_{t}
\end{bmatrix} + 
\begin{bmatrix}
\vec{s}^{x}_{t} \\ \vec{s}^{\hat{x}}_{t}
\end{bmatrix} \enspace . \\
&\nabla_{\delta \vec{\tilde{x}} \delta \vec{\tilde{x}}} \Psi = 
\begin{bmatrix}
\vec{S}^{x}_{t}		 	& \vec{S}^{x \hat{x}}_{t} \\
\vec{S}^{\hat{x} x}_{t} 	& \vec{S}^{\hat{x}}_{t}
\end{bmatrix} \enspace . 
\end{align*}

The right super-scripts $x$ and $\hat{x}$ for $\vec{S}$ and $\vec{s}$ denote that they are sub-blocks that multiply $x$ and $\hat{x}$, respectively. Under the assumed linear dynamics and quadratic cost and value function, the HJB eq. can be written as follows. The LHS corresponds to the time derivative of the value function and is given by
\begin{align}
-& \frac{1}{2} \delta \vec{x}_{t}^{T} \vec{\dot{S}}^{x}_{t} \delta \vec{x}_{t} - \frac{1}{2} \delta \vec{\hat{x}}^{T}_{t} \vec{\dot{S}}^{\hat{x}}_{t} \delta \vec{\hat{x}}_{t} - \delta \vec{x}^{T}_{t} \vec{\dot{S}}^{x \hat{x}}_{t} \delta \vec{\hat{x}}_{t} - \delta \vec{x}^{T}_{t} \vec{\dot{s}}^{x}_{t} - \delta \vec{\hat{x}}^{T}_{t} \vec{\dot{s}}^{\hat{x}}_{t} - \dot{s}_{t} \enspace ,  \label{LHS_CT}
\end{align}
and the RHS corresponds to the following minimization (where for presentation clarity, we call $\alpha_t = \vec{C}_{t} \Omega_{t} \vec{C}_{t}^{T}$ and $\beta_t = \vec{K}_{t} \vec{D}_{t} \Gamma_{t} \vec{D}_{t}^{T} \vec{K}_{t}^{T}$):
\begin{align}
= \min_{\delta \vec{u}_{t}}\ \bigg\{ & q_{t} + \vec{q}^{T}_{t} \delta \vec{x}_{t} + \vec{r}^{T}_{t} \delta \vec{u}_{t} + \frac{1}{2} \delta \vec{x}^{T}_{t}  \vec{Q}_{t} \delta \vec{x}_{t} + \delta \vec{x}^{T}_{t}  \vec{P}_{t} \delta \vec{u}_{t} + \frac{1}{2} \delta \vec{u}^{T}_{t}  \vec{R}_{t} \delta \vec{u}_{t} + \nonumber \\
& (\vec{S}^{x}_{t} \delta \vec{x}_{t} + \vec{S}^{x \hat{x}}_{t} \delta \vec{\hat{x}}_{t} + \vec{s}^{x}_{t})^{T} (\vec{A}_{t} {\delta \vec{x}_{t}} + \vec{B}_{t} {\delta \vec{u}_{t}}) + (\vec{S}^{\hat{x} x}_{t} \delta \vec{x}_{t} + \vec{S}^{\hat{x}}_{t} \delta \vec{\hat{x}}_{t} + \nonumber \\
&\vec{s}^{\hat{x}}_{t})^{T} (\vec{A}_{t} {\delta \vec{\hat{x}}_{t}} + \vec{B}_{t} {\delta \vec{u}_{t}} + \vec{K}_{t} \vec{F}_{t}(\delta \vec{x}_{t} - \delta \vec{\hat{x}}_{t})) + \frac{\sigma}{2} (\vec{S}^{x}_{t} \delta \vec{x}_{t} + \vec{S}^{x \hat{x}}_{t} \delta \vec{\hat{x}}_{t} \nonumber \\
&+ \vec{s}^{x}_{t})^{T} \alpha_t (\vec{S}^{x}_{t} \delta \vec{x}_{t} + \vec{S}^{x \hat{x}}_{t} \delta \vec{\hat{x}}_{t} + \vec{s}^{x}_{t}) + \frac{\sigma}{2} (\vec{S}^{\hat{x} x}_{t} \delta \vec{x}_{t} + \vec{S}^{\hat{x}}_{t} \delta \vec{\hat{x}}_{t} +  \nonumber \\
&\vec{s}^{\hat{x}}_{t})^{T} \beta_t (\vec{S}^{\hat{x} x}_{t} \delta \vec{x}_{t} + \vec{S}^{\hat{x}}_{t} \delta \vec{\hat{x}}_{t} + \vec{s}^{\hat{x}}_{t}) + \frac{1}{2} \text{Tr} \left( \vec{S}^{x}_{t} \alpha_t \right) + \frac{1}{2} \text{Tr} \left( \vec{S}^{\hat{x}}_{t} \beta_t \right) \bigg\} \enspace . \label{RHS_HJB_CT}
\end{align}
To perform the minimization of the RHS, we analyze its control dependent terms, corresponding to the part of the cost to go that is control dependent:
\begin{align}
V_{\delta \vec{u}_{t}} = &\frac{1}{2} \delta \vec{u}_{t}^{T}  \underbrace{\vec{R}_{t}}_{\vec{H}_{t}} \delta \vec{u}_{t} + \delta \vec{u}_{t}^{T} \big( \underbrace{\vec{r}_{t} + \vec{B}_{t}^{T} \left( \vec{s}^{x}_{t} + \vec{s}^{\hat{x}}_{t} \right)}_{\vec{g}_{t}} + \nonumber \\
& \underbrace{\left( \vec{P}_{t}^{T} + \vec{B}_{t}^{T} \left( \vec{S}^{x}_{t} + \vec{S}^{\hat{x} x}_{t} \right) \right)}_{\vec{G}^{x}_{t}} \delta \vec{x}_{t} + \underbrace{\vec{B}_{t}^{T} \left( \vec{S}^{x \hat{x}}_{t} + \vec{S}^{\hat{x}}_{t} \right)}_{\vec{G}^{\hat{x}}_{t}} \delta \vec{\hat{x}}_{t} \big) \enspace . \label{control_dep_value_function_terms_CT}
\end{align}

The above expression is quadratic in $\delta \vec{u}_{t}$ and is easy to minimize. However, the minimum is a functional not only of $\delta \vec{\hat{x}}_{t}$, but also of $\delta \vec{x}_{t}$. Here, we use the assumption that we do not have access to full state information, only a statistical description of it, given by the state estimate. Therefore, in order to perform the minimization, we take an expectation of $V_{\delta \vec{u}_{t}}$ over $\delta \vec{x}_{t}$ conditioned on $\delta \vec{\hat{x}}_{t}$
\begin{align}
\mathbb{E}_{\delta \mathbf{x}_{t} | \delta \mathbf{\hat{x}}_{t}} \left[ V_{\delta \vec{u}_{t}} \right]  = \frac{1}{2} \delta \vec{u}^{T}_{t}  \vec{H}_{t} \delta \vec{u}_{t} + \delta \vec{u}_{t}^{T} ({\vec{g}_{t}} + (\vec{G}^{x}_{t}+\vec{G}^{\hat{x}}_{t})\delta \vec{\hat{x}}_{t}) \enspace . \nonumber 
\end{align}

This means that the cost of uncertainty due to measurement noise, considers only the effects of mean and variance of the measurement (captured by the EKF) when evaluating noise effects on the statistical properties of the performance criteria. Consequently, the risk-sensitive control law, considers only as cost of measurement uncertainty the one that can be computed by means of the state estimate, in other words, the one that can be extracted from using mean and variance of the state estimate and neglecting higher order terms.

From the above expression, the minimizer can be analytically computed. In case of control constraints, a quadratic program can be used to solve for the constrained minimizer \cite{control_limited_DDP}. In both cases, the minimizer is an affine functional of the state-estimate. For the unconstrained case, it is given by
\begin{align}
\delta \vec{u}_{t} = \vec{l}_{t} + \vec{L}_{t} \delta \vec{\hat{x}}_{t} = - \vec{H}_{t}^{-1} {\vec{g}_{t}} - \vec{H}_{t}^{-1} (\vec{G}^{x}_{t}+\vec{G}^{\hat{x}}_{t}) \delta \vec{\hat{x}}_{t} \enspace . \label{minimizer_CT}
\end{align}
$V_{\delta \vec{u}_{t}}$ can then be written in terms of the optimal control as
\begin{align}
V_{\delta \vec{u}_{t}^{*}} = & \frac{1}{2} \delta \vec{\hat{x}}_{t}^{T} \left( (\vec{G}^{x}_{t})^{T} \vec{H}_{t}^{-1} \vec{G}^{x}_{t} - (\vec{G}^{\hat{x}}_{t})^{T} \vec{H}_{t}^{-1} \vec{G}^{\hat{x}}_{t}  \right) \delta \vec{\hat{x}}_{t} - \delta \vec{x}_{t}^{T} (\vec{G}^{x}_{t})^{T} \vec{H}_{t}^{-1} (\vec{G}^{x}_{t} + \nonumber \\
& \vec{G}^{\hat{x}}_{t}) \delta \vec{\hat{x}}_{t} - \frac{1}{2} \vec{g}_{t}^{T} \vec{H}_{t}^{-1} \vec{g}_{t} - \delta \vec{x}_{t}^{T} (\vec{G}^{x}_{t})^{T} \vec{H}_{t}^{-1} \vec{g}_{t} - \delta \vec{\hat{x}}_{t}^{T} (\vec{G}^{\hat{x}}_{t})^{T} \vec{H}_{t}^{-1} \vec{g}_{t} \enspace .  \label{control_dependent_terms_minimizer_CT}
\end{align}

The negative coefficients in the terms of $V_{\delta \vec{u}_{t}^{*}}$ are the benefit of control at reducing the cost. It should be noted that even setting measurement noise to zero does not give a control law equivalent to what was found in \cite{iLEG}. It should be clear from \eqref{minimizer_CT} that mathematically they are not the same. However, it is worth pointing out that, \cite{iLEG} considers neither measurement noise, nor the combined effect of process and measurement noise over optimal controls. Here, we do, and setting measurement noise to zero has the specific meaning that we are absolutely sure about our state, and because of it, this control law allows the use of more control authority. In the presence of measurement noise, our control law has more conservative gains than \cite{iLEG}, in order to remain compliant enough for the measurement noise level. Writing these terms back into the RHS of the HJB, we can drop the minimization and verify that the quadratic Ansatz for the value function remains quadratic and is therefore valid. Finally, matching terms in LHS and RHS of the HJB eq., we write the backward pass recursion eqns. as:
%
\begin{align}
- \vec{\dot{S}}^{x}_{t} = &\vec{Q}_{t} + \vec{A}_{t}^{T} \vec{S}_{t}^{x} + (\vec{S}_{t}^{x})^{T}\vec{A}_{t} + \vec{S}^{x \hat{x}}_{t} \vec{K}_{t} \vec{F}_{t} +  \vec{F}_{t}^{T} \vec{K}_{t}^{T} (\vec{S}_{t}^{x \hat{x}})^{T} + \nonumber \\
&\sigma (\vec{S}_{t}^{x})^{T} \alpha_t \vec{S}_{t}^{x} + \sigma \vec{S}_{t}^{x \hat{x}} \beta_t (\vec{S}_{t}^{x \hat{x}})^{T} \enspace . \nonumber \\
- \vec{\dot{S}}^{\hat{x}}_{t} = & (\vec{A}_{t}-\vec{K}_{t} \vec{F}_{t})^{T} \vec{S}_{t}^{\hat{x}} + (\vec{S}_{t}^{\hat{x}})^{T}(\vec{A}_{t}-\vec{K}_{t} \vec{F}_{t}) + (\vec{G}^{x}_{t})^{T} \vec{H}_{t}^{-1} \vec{G}^{x}_{t} - \nonumber \\
&(\vec{G}^{\hat{x}}_{t})^{T} \vec{H}_{t}^{-1} \vec{G}^{\hat{x}}_{t} + \sigma (\vec{S}_{t}^{x \hat{x}})^{T} \alpha_t \vec{S}_{t}^{x \hat{x}} + \sigma (\vec{S}_{t}^{\hat{x}})^{T} \beta_t \vec{S}_{t}^{\hat{x}} \enspace . \nonumber \\
- \vec{\dot{S}}^{x \hat{x}}_{t} = &\vec{A}_{t}^{T} \vec{S}_{t}^{x \hat{x}} + \vec{S}_{t}^{x \hat{x}} (\vec{A}_{t}-\vec{K}_{t} \vec{F}_{t}) + \vec{F}_{t}^{T} \vec{K}_{t}^{T} \vec{S}_{t}^{\hat{x}} - (\vec{G}^{x}_{t})^{T} \vec{H}_{t}^{-1} (\vec{G}^{x}_{t}+\vec{G}^{\hat{x}}_{t}) + \nonumber \\
&\sigma (\vec{S}_{t}^{x})^{T} \alpha_t \vec{S}_{t}^{x \hat{x}} + \sigma \vec{S}_{t}^{x \hat{x}} \beta_t \vec{S}_{t}^{\hat{x}} \enspace . \nonumber \\
- \vec{\dot{s}}^{x}_{t} = &\vec{q}_{t} + \vec{A}_{t}^{T} \vec{s}_{t}^{x} + \vec{F}_{t}^{T} \vec{K}_{t}^{T} \vec{s}_{t}^{\hat{x}} - (\vec{G}^{x}_{t})^{T} \vec{H}_{t}^{-1} \vec{g}_{t} + \sigma (\vec{S}_{t}^{x})^{T} \alpha_t \vec{s}_{t}^{x} + \sigma \vec{S}_{t}^{x \hat{x}} \beta_t \vec{s}_{t}^{\hat{x}} \enspace . \nonumber \\
- \vec{\dot{s}}^{\hat{x}}_{t} = &(\vec{A}_{t}-\vec{K}_{t} \vec{F}_{t})^{T} \vec{s}_{t}^{\hat{x}} - (\vec{G}^{\hat{x}}_{t})^{T} \vec{H}_{t}^{-1} \vec{g}_{t} + \sigma (\vec{S}_{t}^{x \hat{x}})^{T} \alpha_t \vec{s}_{t}^{x} + \sigma (\vec{S}_{t}^{\hat{x}})^{T} \beta_t \vec{s}_{t}^{\hat{x}} \enspace . \nonumber \\
- \vec{\dot{s}}_{t} = & q_{t} - \frac{1}{2} \vec{g}_{t}^{T} \vec{H}_{t}^{-1} \vec{g}_{t} + \frac{1}{2} \text{Tr} \left( \vec{S}_{t}^{x} \alpha_t \right) + \frac{1}{2} \text{Tr} \left( \vec{S}_{t}^{\hat{x}} \beta_t \right) + \nonumber \\
&\frac{\sigma}{2} (\vec{s}_{t}^{x})^{T} \alpha_t \vec{s}_{t}^{x} + \frac{\sigma}{2} (\vec{s}_{t}^{\hat{x}})^{T} \beta_t \vec{s}_{t}^{\hat{x}} \enspace .  \label{backward_recursion_CT}
\end{align}
The integration runs backward in time with $\vec{S}^{x}_{t} = \vec{Q}_{f}$, $\vec{S}^{\hat{x}}_{t} = 0$, $\vec{S}^{x \hat{x}}_{t} = 0$, $\vec{s}^{x}_{t} = \vec{q}_{f}$, $\vec{s}^{\hat{x}}_{t} = 0$ and $s_{t} = q_{f}$. Despite being long, it is a very simple to implement solution, similar to any other LQR-style recursion.
\begin{remark}
The effects of process and measurement noise appear in pairs due to the fact that we assumed their Brownian motions to be uncorrelated (see $\vec{g}(t)$ in \eqref{enlarged_dynamics_system_CT}). However, their combined effect is not just as having higher process noise. Estimation couples their effects, and this can be seen in the recursion equation, where we do not only have costs for the state and its estimate $\vec{S}_{t}^{x}$ and $\vec{S}_{t}^{\hat{x}}$, but also the coupling cost $\vec{S}_{t}^{x \hat{x}}$; whose products with the covariances of process noise and estimation error determine how process noise and measurement uncertainty affect the value function and therefore the control law. 
\end{remark}
%

\vspace{-0.4cm}
\section{Experimental Results}
\vspace{-0.2cm}
In this section, we use the control algorithm on a 2-DOF manipulator on 2 different tasks: a viapoint task and a contact task. This setup allows us to analyze in a simple setting the important properties of the algorithm. The equations of motion are given by
\begin{align}
\vec{H}(\vec{q}) \ddot{\vec{q}} + \vec{C}(\vec{q},\dot{\vec{q}}) = \boldsymbol{\tau} + \vec{J}(\vec{q})^{T} \vec{\lambda} \enspace . \label{eq:man_dyn}
\end{align}

The vector $\vec{q} = [q_{1},\ q_{2}]^{T}$ contains the joints positions. $\vec{H}(\vec{q})$ is the inertia matrix, $\vec{C}(\vec{q},\dot{\vec{q}})$ the vector of Coriolis and centrifugal forces, $\vec{J}(\vec{q})$ is the end-effector Jacobian, $\vec{\lambda} \in \bbbr^{2}$ the external forces and $\vec{\tau} \in \bbbr^{2}$ the input torques. The system dynamics can be easily written in the form given by \eqref{eqn_dyn},
with additive process noise $d\omega$ and state $\vec{x} = [\vec{q}^{T},\ \dot{\vec{q}}^{T}]^{T}$. The measurement model can also easily be written in the form given by \eqref{eqn_mea} ($d\vec{y}=d\vec{x}+d\gamma$), with Brownian motion $\gamma$ with variance $\Gamma dt$.
\vspace{-0.6cm}
\begin{figure}[H]
	\centering
	\includegraphics[width=45mm]{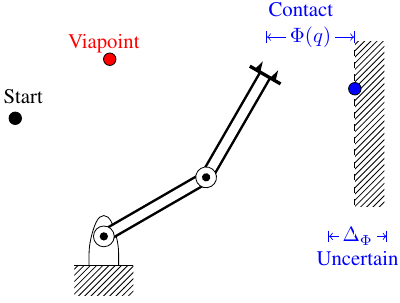}
   	\vspace{-0.3cm}
    \caption{Schematic: pass through a viapoint and establish contact with a wall with uncertain location $\Phi(q)+\Delta_{\Phi}$.}
    \label{fig_twoarm}
   	\vspace{-0.7cm}
\end{figure}
\subsection{Experiment 1: Process Noise vs. Measurement Uncertainty}
We compare the effect of process and measurement noise in the control law in a motion task between two points with two viapoints. The objective function
\vspace{-0.1cm}
\begin{align*}
\mathcal{J} = &\sum_{0}^{t_{f}}c_{u} \vec{\tau}^{T}\vec{\tau} + \sum_{i=1}^{N_{via}} c_{i} \log(\cosh(||\vec{x}-\vec{x}_{i}||_{2})) + c_{t_{f}} \log(\cosh(||\vec{x}-\vec{x}_{t_{f}}||_{2})) \nonumber
\end{align*}
measures task performance. $\vec{x}$, $\vec{x}_{i}$, $\vec{x}_{t_{f}} \in \bbbr^{4}$ are current, viapoints and final desired end-effector positions and velocities, respectively. $c_{u}$, $c_{i}$, $c_{t_{f}}$ are cost weights. The nonlinear cost $\log(\cosh(\cdot))$ is a soft absolute value to demonstrate that general nonlinear costs functions can be used.

\begin{figure}[h]
	\centering
	\includegraphics[width=86mm]{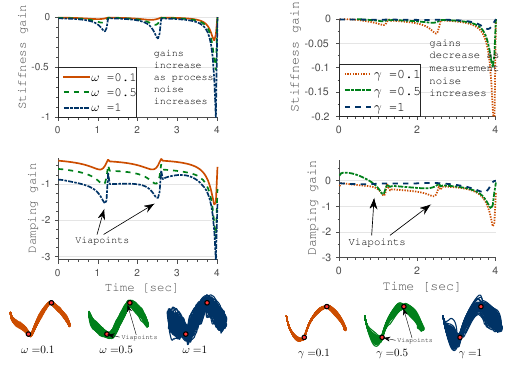}
	\caption{Comparison between process noise and measurement uncertainty. The control gains for various level of noise are shown in the upper graphs. Sample trajectories are shown in the lower graphs for both varying process noise (left) and measurement uncertainty (right). The red dots represent the viapoints. In all the experiments the risk sensitive parameter $\sigma = 2.5$. Stiffness gains are normalized to 1 corresponding to 100 N/m.}
	\label{fig_experiment001}
    \vspace{-0.6cm}
\end{figure}

We first evaluate the effects of increasing process noise under no measurement uncertainty (Fig. \ref{fig_experiment001} - left). Feedback gains for the motion task under several noise intensities are shown. In general, they are higher for regulating behavior at the viapoints and goal position. As process noise increases, the cost of uncertainty does too, because we might miss the viapoints or the goal due to disturbances. This can be seen in sample trajectories, where the variance of the trajectories due to noise has increased. In this case, the trade-off between cost of uncertainty and control-effort involves feedback gains proportional to the process noise, {\bfseries the higher the process noise, the higher the feedback gains}.

In a second set of simulations, we test the effect of increasing measurement uncertainty under no process noise (Fig. \ref{fig_experiment001} - right). Feedback gains and sample trajectories for different values of measurement noise are shown. Feedback gains are also higher near viapoints and goal position, and sample trajectories are similar to the ones with process noise. The big difference is that the optimal control solution for this case is to trust feedback proportionally to the information content of the measurements, namely, {\bfseries the higher the measurement uncertainty, the lower the feedback gains}. It shows how under low measurement noise, feedback control with higher gains is possible and optimal. Under high measurement noise, lower impedance is better. We note that during the evaluation of the controller online estimation is used as it achieves better performance than using the precomputed sequence of estimation gains. In these experiments, we kept the risk sensitive parameter constant as it is not the focus of this paper (see for example \cite{iLEG}). However, the effects of process and measurement noise are qualitatively similar for all allowed values of $\sigma$ (data not shown).
\vspace{-0.3cm}
\subsection{Experiment 2: Establishing Contact with the Environment}
\begin{figure}[h]
\vspace{-0.4cm}
	\centering
	\includegraphics[width=100mm]{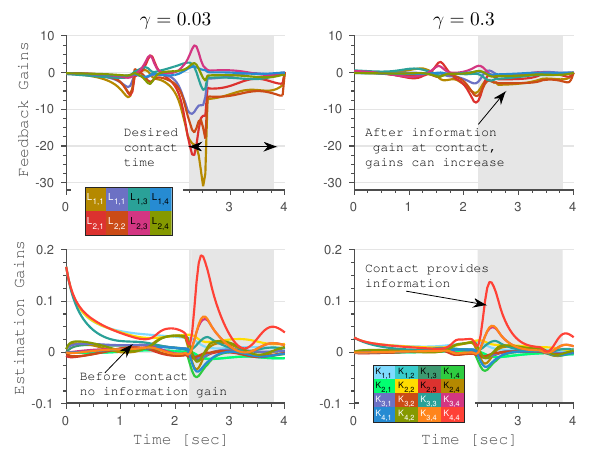}
	\caption{Feedback and Estimation Gains for a motion-force task for two different values of measurement noise $\gamma$, that encodes the uncertainty in the distance to the contact.}
	\label{fig_exp_003a}
	\vspace{-0.7cm}
\end{figure}

\begin{figure}[h]
	\centering
	\includegraphics[width=100mm]{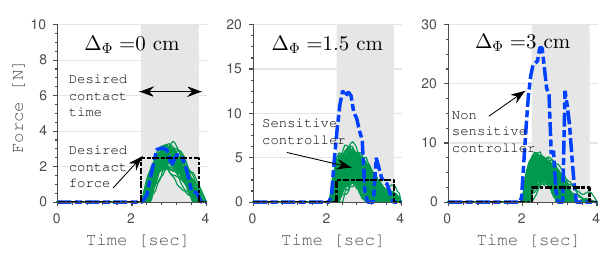}
	\caption{Contact forces given a perturbation $\Delta \Phi$ at the contact location. Black dashed lines show reference desired forces and dashed blue lines show the interaction profile of a typical optimal feedback controller that	does not take into account measurement uncertainty. The controller sensitive to measurement uncertainty (green lines) robustly deal with the uncertain dynamic interaction.}
	\label{fig_exp_003b}
	\vspace{-0.6cm}
\end{figure}

In this experiment, the robot needs to pass through a viapoint and then make contact with a wall at an uncertain location ($\Delta_\Phi$), as shown in Fig. \ref{fig_twoarm}. While simple enough to be carefully analyzed, the experiment addresses the role of measurement uncertainty when interacting with an uncertain environment, which is important for manipulation and locomotion tasks. Performance is measured by
\vspace{-0.2cm}
\begin{align*}
\mathcal{J} = &\sum_{0}^{t_{f}} c_{\scriptsize \mbox{u}} \vec{\tau}^{T}\vec{\tau} + c_{\scriptsize \mbox{via}} \log (\cosh(||\vec{x}_{t_{\scriptsize \mbox{via}}}-\vec{x}_{\scriptsize \mbox{via}}||_{2})) + \nonumber \\ 
& \sum_{t_{\scriptsize \mbox{cnt}_{0}}}^{t_{\scriptsize \mbox{cnt}_{f}}} c_{\scriptsize \mbox{cnt}} \log(\cosh(\Phi(\vec{q})) \cosh(||\vec{\lambda}-\vec{\lambda}_{\scriptsize \mbox{des}}||_{2})) \enspace . \nonumber
\end{align*}
\noindent $\vec{\lambda}_{\scriptsize \mbox{des}} \in \bbbr^{2}$ is the desired contact force at contact; $c_{\scriptsize \mbox{cnt}}$, $c_{\scriptsize \mbox{via}}$, $c_{\scriptsize \mbox{u}}$ are cost weights. This cost rewards low torques, passing a viapoint $x_{\scriptsize \mbox{via}}$ at time $t_{\scriptsize \mbox{via}}$, making contact $\Phi(\vec{q})=0$ and exerting a desired force from $t_{\scriptsize \mbox{cnt}_{0}}$ to $t_{\scriptsize \mbox{cnt}_{f}}$ (shown as shaded areas in Figs. \ref{fig_exp_003a}-\ref{fig_exp_003b}). The external force $\vec{\lambda}$ is modeled as a stiff spring and is part of the dynamic model such that its effect is known to the optimizer. 
Given that its value depends on the uncertain position of the wall, it is also an uncertain variable.
There are two possible ways to encode the uncertainty in the distance to the contact $\Phi(\vec{q})$. On the one hand, it is a function of the joint positions $q$, and therefore, we could model measurement noise directly into these components. The other alternative is to add a new state $x^{'}$ to the state vector $\vec{x}$. This new state would be defined as the distance to the contact $\Phi(\vec{q})$. For the dynamic model, we need its derivative, which is given by $\nabla_{\vec{q}} \Phi(\vec{q})^{T}\vec{\dot{q}}$, and in the measurement model we model directly our uncertainty in the value of $\Phi(\vec{q})$.

Fig. \ref{fig_exp_003a} shows feedback and estimation gains for two measurement noise values $\gamma$ that encode uncertainty in the state (distance to contact). Feedback gains show two peaks around $1$ and $2.2\,\mathrm{sec}$, when passing the viapoint and when contact happens. Feedback gains for $\gamma = 0.03$ are higher than for $\gamma = 0.3$, where control is more cautious. Estimation gains are qualitatively similar. Interestingly, passing the viapoint does not affect them but when the contact is expected, they are higher because contact provides location information. As we expected, under increasing measurement noise, feedback gains decrease, which allows us to have a compliant interaction in the presence of measurement uncertainty.

Fig. \ref{fig_exp_003b} shows force profiles of contact interactions with the wall. Black dashed lines are the reference forces. Dashed blue lines show the interaction force profiles using a controller not sensitive to measurement uncertainty, optimized for process noise ($\omega=0.2$) but very low measurement uncertainty ($\gamma = 0.003$). The distribution of force profiles under the sensitive control law and the stochastic dynamics is shown in green. It was optimized for process noise ($\omega=0.2$) and measurement uncertainty ($\gamma = 0.3$). We see that with the controller using measurement uncertainty when contact happens before it was expected ($\Delta_{\Phi} = 1.5$ or $\Delta_{\Phi} = 3.0$ cm), forces are higher than the reference, but the interaction is not as aggressive as it would be with the higher feedback gains of a usual non-sensitive optimal controller. In the case of the controller sensitive only to process noise, since the feedback gains are higher we see much higher contact forces (blue lines) and even a loss of contact (Fig. \ref{fig_exp_003b} - right).

These results illustrate a behavior relevant for robotic applications involving contact interactions: policies sensitive to measurement uncertainty lead to low impedance behavior in face of too high uncertainty. In a receding horizon setting, the impedance behavior would then be adapted as the robot gains more information about the state of the environment (e.g. after making a contact). While the execution does not exploit sensed contact forces, which could improve further the dynamic interaction, it is still able to find a feedback control policy that can safely interact with the environment, despite uncertainty in the position of the wall. This example illustrates that taking into account measurement uncertainty in the control law can lead to more robust behaviors for contact interaction, that cannot be achieved with an approach taking only into account process noise. It is worth noting that our noise description is not limited to state transition due to actuators noise or sensor measurements noise, but can include uncertainty on process and measurement models, which are only an approximation of the true underlying dynamics.
\vspace{-0.3cm}
\section{Discussion}\label{sec:discussion}
\vspace{-0.2cm}
In our experiments, we have seen that process noise is fundamentally different from measurement noise. While the first one is a dynamics disturbance that requires control using high feedback gains; the second one represents uncertainty in the state information, and requires compliance proportional to the uncertainty to dynamically interact with the world given our limited knowledge of it.

The fundamental difference between process and measurement noise effects on the control law comes from the cost they penalize. Cost of uncertainty due to process noise increases with terms of the form $\vec{C}_{t} \Omega_{t} \vec{C}_{t}^{T}$. If there is no control action, the process noise increases the cost. Therefore, regulation with high gains is optimal. For measurement noise, cost increases with terms $\vec{K}_{t} \vec{D}_{t} \Gamma_{t} \vec{D}_{t}^{T} \vec{K}_{t}^{T}$ and estimation gains are inversely proportional to measurement noise. Therefore, not using informative measurements is costly and requires high feedback gains. For poorly-informative measurements, we incur very low cost and control with lower gains is optimal. This behavior can be exploited in robotic tasks with dynamic interactions. For example when making a contact, behaving compliant under poor contact-information is robust. Once the contact is established and position certainty is higher, feedback gains would then be increased. In a receding horizon setup, gain in information about the current state of the world after contact would allow to online adapt the feedback policy, which would improve the performance of the controller compared to only executing a plan, as shown in Fig. \ref{fig_exp_003b}.

In this work, we have looked into a problem with simple geometry and unidirectional contact interaction. Besides, the uncertainty in the distance to contact has been modelled as a Gaussian distribution. While the first two assumptions, using a point contact model, are common and therefore transferable to more general cases of multi-body systems interacting with a more geometrically rich environment, it is not clear if the same holds for the uncertainty model as a Gaussian distribution. Despite of this fact, the value of this work resides in the insight provided about what is important to consider for controlling the dynamic interaction of a robot with its environment.

From a computational point of view, the algorithm should scale to more complex systems. We can approximate the complexity of a call to the dynamics with its heaviest computation (factorization and back-substitution of $H(q)$) as roughly $O(n^{3})$, $n$ being the number of states. The most expensive computation is that of first derivatives $O(N n^{4})$, $N$ being the number of timesteps in the horizon. This is in the same order of complexity as other iterative approaches that show very good performance on more complicated robotic tasks \cite{TassaMPC},\cite{control_limited_DDP}, although those examples did not exploit measurement uncertainty for control. While our approach requires using twice the number of actual states, which increases the solving time a small amount, this should still be fine given the impressive results of recent papers on high dimensional robotics problems with contacts \cite{DBLP:conf/icml/LevineK13}.
\vspace{-0.35cm}
\section{Conclusion}\label{sec:conclusion}
\vspace{-0.35cm}
We have presented an iterative algorithm for finding locally-optimal feedback controllers for nonlinear systems with additive measurement uncertainty. In particular we showed that measurement uncertainty leads to very different behaviors than process noise and it can be exploited to create low impedance behaviors in uncertain environments (e.g. during contact interaction). This opens the possibility for planning and controlling contact interactions robustly based on controllers sensitive to measurement noise. In a receding horizon setting, it could be possible to regulate impedance in a meaningful way depending on the current uncertainty about the environment.

\vspace{0.2cm} \textbf{Acknowledgments.} This research was mainly supported by the Max-Planck-Society and the European Research Council under the European Union’s Horizon 2020 research and innovation programme (grant No 637935). It was also supported by National Science Foundation grants IIS-1205249, IIS-1017134, EECS-0926052, the Office of Naval Research, the Okawa Foundation, and the Max-Planck ETH Center for Learning Systems.

\vspace{-0.1cm}
\bibliographystyle{unsrt}
\vspace{-0.2cm}
\bibliography{references}

\end{document}